\documentclass[conference]{IEEEtran}
\IEEEoverridecommandlockouts
\usepackage{cite}
\usepackage{amsmath,amssymb,amsfonts}
\usepackage{algorithmic}
\usepackage{graphicx}
\usepackage{textcomp}
\usepackage{xcolor}
\usepackage{wrapfig}
\usepackage{textcomp}
\usepackage{xcolor}
\usepackage{soul}
\usepackage{braket}
\usepackage{algorithmic}
\usepackage[ruled,vlined, linesnumbered]{algorithm2e}
\usepackage{graphicx}
\usepackage{cite}
\usepackage{amsmath,amssymb,amsfonts}
\usepackage{algorithmic}
\usepackage{graphicx}
\usepackage{textcomp}
\usepackage{xcolor}
\usepackage{wrapfig}
\usepackage{textcomp}
\usepackage{xcolor}
\usepackage{soul}
\usepackage{braket}
\usepackage{algorithmic}
\usepackage[ruled,vlined, linesnumbered]{algorithm2e}
\usepackage{graphicx}
\usepackage{cite}
\usepackage{amsmath,amssymb,amsfonts}
\usepackage{algorithmic}
\usepackage{graphicx}
\usepackage{textcomp}
\usepackage{xcolor}
\def\BibTeX{{\rm B\kern-.05em{\sc i\kern-.025em b}\kern-.08em
    T\kern-.1667em\lower.7ex\hbox{E}\kern-.125emX}}
\begin{document}

\title{Stealthy SWAPs: Adversarial SWAP Injection in Multi-Tenant Quantum Computing\\
 }

\author{\IEEEauthorblockN{Suryansh Upadhyay}
\IEEEauthorblockA{\textit{The Pennsylvania State University} \\
\textit{University Park, PA, USA} \\
sju5079@psu.edu}
\and
\IEEEauthorblockN{Swaroop Ghosh}
\IEEEauthorblockA{\textit{The Pennsylvania State University} \\
\textit{University Park, PA, USA} \\
szg212@psu.edu}
}

\maketitle

\begin{abstract}
Quantum computing (QC) holds tremendous promise in revolutionizing problem-solving across various domains. It has been suggested in literature that 50+ qubits are sufficient to achieve quantum advantage (i.e., to surpass supercomputers in solving certain class of optimization problems).
The hardware size of existing Noisy Intermediate-Scale Quantum (NISQ) computers have been ever increasing over the years.
Therefore, Multi-tenant computing (MTC) has emerged as a potential solution for efficient hardware utilization, enabling shared resource access among multiple quantum programs. However, MTC can also bring new security concerns. This paper proposes one such threat for MTC in superconducting quantum hardware i.e., adversarial SWAP gate injection in victim's program during compilation for MTC. We present a representative scheduler designed for optimal resource allocation. To demonstrate the impact of this attack model, we conduct a detailed case study using a sample scheduler. Exhaustive experiments on circuits with varying depths and qubits offer valuable insights into the repercussions of these attacks. We report a max of $\approx$55\% and a median increase of $\approx$25\% in SWAP overhead. As a countermeasure, we also propose a sample machine learning model for detecting any abnormal user behavior and priority adjustment.
\end{abstract}


\section{Introduction}

Quantum computing (QC) has garnered considerable attention due to its potential to revolutionize problem-solving across various domains. Leveraging quantum-mechanical phenomena, such as superposition and entanglement, quantum computers offer exponential speed-ups over classical counterparts in solving combinatorial problems. The applications of quantum computing extend to machine learning\cite{b1}, security\cite{b2}, drug discovery\cite{b3}, and optimization\cite{b4}, making it a highly sought-after technology in the scientific and commercial communities. However, practical implementation of quantum computing faces formidable challenges, including qubit decoherence, measurement errors, gate errors, and temporal variations. Quantum error correction (QEC) codes \cite{b5} provide a solution for reliable quantum operations, but their current resource-intensive requirements render them impractical for widespread use in the near future. The emergence of Noisy Intermediate-Scale Quantum (NISQ) computers, which have a limited number of qubits and operate in the presence of noise, offer a potential solution to important problems such as discrete optimization and quantum chemical simulations. 

Quantum computers are typically accessed via cloud services, which offer users convenience and scalability. However, the increasing number of quantum computer users has resulted in a demand-supply imbalance, leading to significant wait times for accessing quantum computing resources. This situation highlights a pressing concern: how can we utilize quantum hardware more efficiently? To address this challenge, the concept of multi-tenant computing (MTC)  has emerged \cite{b6}\cite{b7}, gaining prominence with the rise in hardware qubit numbers and improved qubit error rates. The MTC paradigm seems to be a necessary solution to optimize quantum hardware utilization and meet the growing demands of quantum computing efficiently.

\begin{figure}
    \centering
    \includegraphics[width= 3.5in]{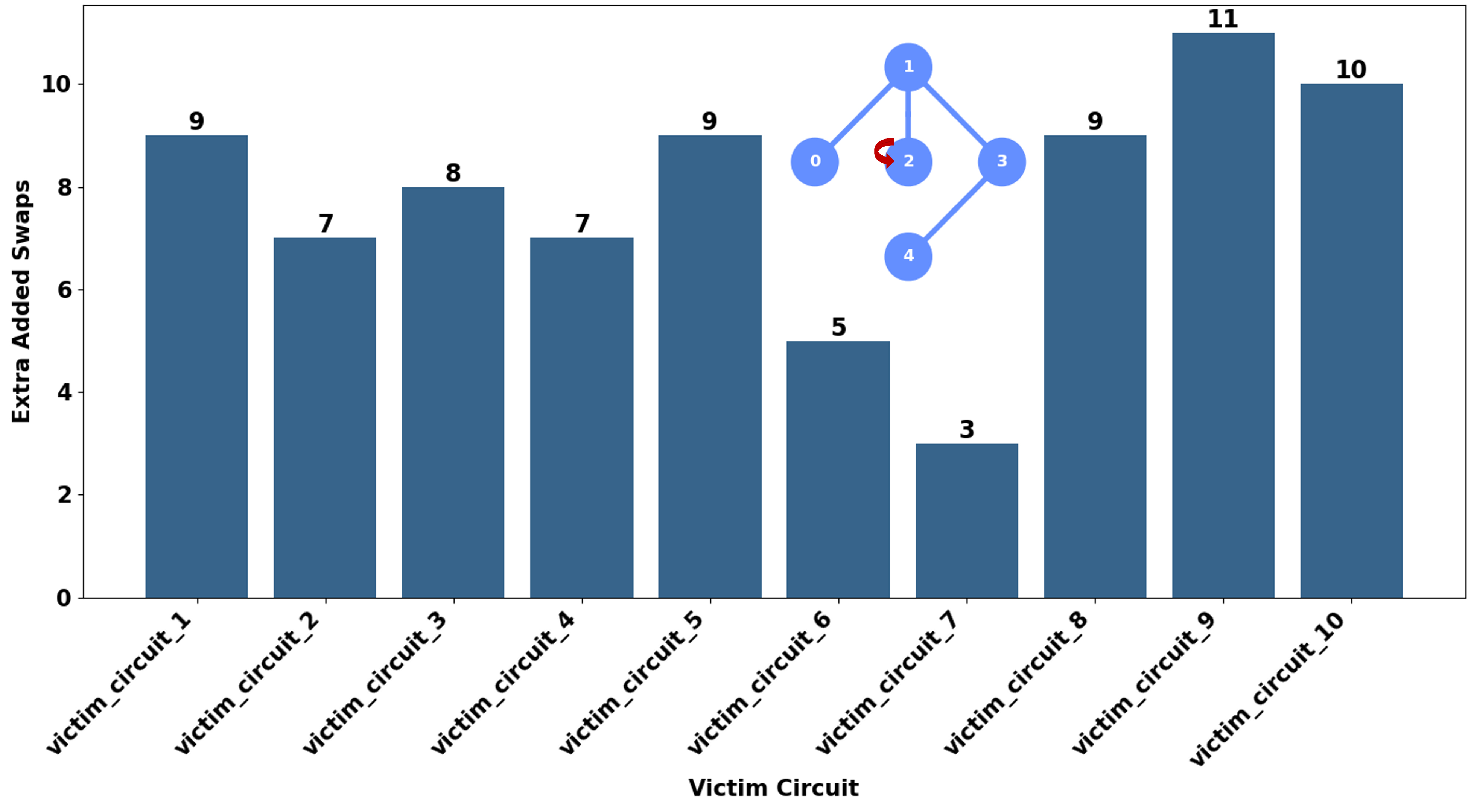}
    \caption{The number of extra SWAPs introduced on the victim circuit when an adversary (1 qubit program) and a victim (4 qubit program) are run on a 5 qubit hardware with the connectivity graph (top left). The graph depicts the results of ten different programs with similar depth but different two-qubit gate permutations. We compare two cases: \textbf{Adversarial attack:} The victim program is allocated qubits 0, 1, 3, and 4, while the adversary program strategically occupies qubit 2. \textbf{Baseline:} The victim program is allocated qubits 0, 1, 2, and 3, while the adversary program is allocated qubit 4.}
    \label{1}
\vspace{-4mm}
\end{figure}

\textbf{Motivation:} In NISQ computers, resource sharing presents significant challenges compared to conventional computers. The limited connectivity between qubits in existing quantum computers necessitates the use of SWAP operations for 2-qubit gates between qubits that are not directly connected. This incurs additional computational costs and introduces error accumulation. In a multi-tenant computing  scenario, adversaries can exploit these connectivity constraints by strategically occupying specific qubits, by exploiting loop holes in scheduler policies. Consequently, victim programs assigned to qubits with poor connectivity experience additional SWAPs. To illustrate this, let's consider an example scenario where we have an adversary with a 1-qubit program and a victim with a 4-qubit program, both running on a 5-qubit hardware with a certain connectivity graph (Fig. \ref{1}). We compare two cases: \textbf{Case-1:} The victim program is allocated qubits 0, 1, 3, and 4, while the adversary program strategically occupies qubit 2. \textbf{Case-2:} The victim program is allocated qubits 0, 1, 2, and 3, while the adversary program is allocated qubit 4. In the first case, the adversary occupies qubit 2, which forces the victim circuit to utilize qubits (0, 1, 3, 4) that are not as densely connected. As a result, the victim program experiences a higher number of SWAPs due to the poor connectivity of its allocated qubits. This negatively impacts its performance and fidelity of computation. Conversely, in the second case, the victim program runs on qubits that are more densely connected, reducing the number of SWAPs required and improving its overall performance.

Understanding and addressing such qubit connectivity-based attacks is crucial for ensuring the reliability and security of quantum computations in MTC environments. This paper focuses on identifying the threats posed by adversary-induced extra SWAPs and their potential impact on the reliability and security of quantum computations.
\emph{To the best of our knowledge, this is the first attempt to delve into the new SWAP-based fault injection threat space tailored specifically for MTC quantum computing.}

\textbf{Contributions:} We present (a) a novel SWAP-based fault injection model, (b) a representative scheduler for MTC, (c) exhaustive experiments using circuits of varying depths and qubits, and (d) countermeasures against such attacks.

\textbf{Paper organization:} Section II provides background information on quantum computing and related work. Section III describes a representative scheduler for MTC. In section IV we elaborate on the proposed threat model. Section V discusses simulations, results, and analysis. In Section VI we discuss countermeasures. Section VII concludes the paper.

\section{Background}


\subsection{Qubits and Quantum gates}

Qubits are similar to classical bits in that they store data through internal states such as $\ket{0}$ and $\ket{1}$. However due to their quantum nature, qubits can exist in a superposition of both $\ket{0}$ and $\ket{1}$. 
Furthermore, qubits can be entangled, which means that the states of multiple qubits become correlated. 
Mathematically, quantum gates are represented using unitary matrices (a matrix U is unitary if U$U^\dagger$ = I, where U$^\dagger$ is the adjoint of matrix U and I is the identity matrix). 

\begin{figure*}
    \centering
    \includegraphics[width= 7.15in]{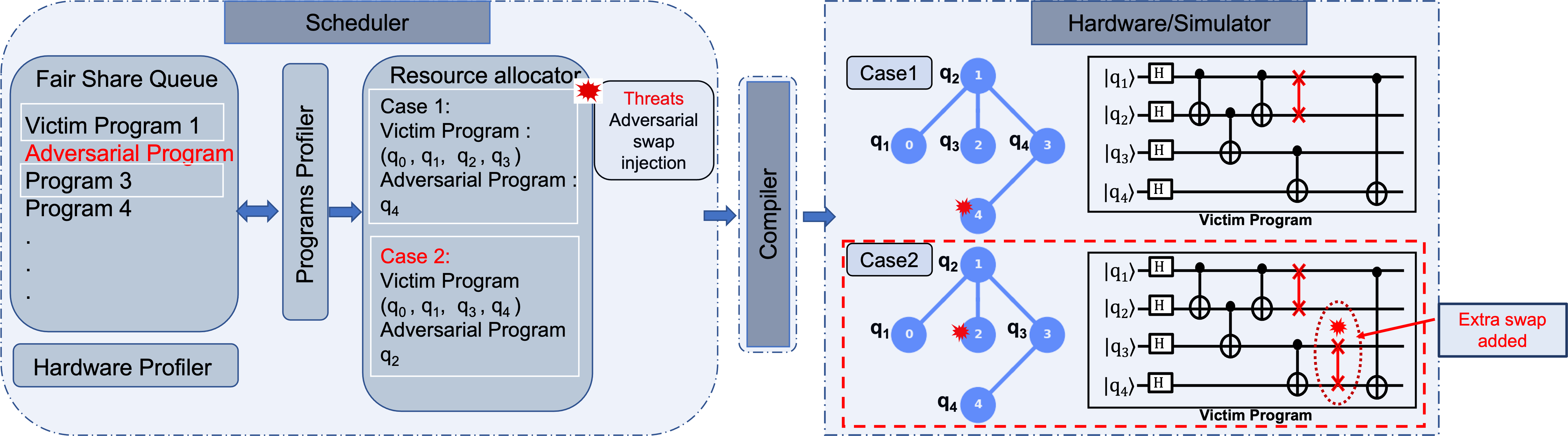}
    \caption{Threat model in MTC: The scheduler determines job execution order and concurrent program allocations on the quantum hardware for a single run in MTC. An adversary can exploit scheduler oversights and quantum hardware connectivity constraints. We compare two cases \textbf{Baseline (Case-1):} The victim program is allocated qubits 0, 1, 2, and 3, while the adversary program is allocated qubit 4. \textbf{Adversarial Attack (Case-2):} The victim program is allocated qubits 0, 1, 3, and 4, while the adversary program strategically occupies qubit 2.  The  extra SWAP introduced on the victim circuit highlights the impact of the attack.}
    \label{2}
\vspace{-4mm}
\end{figure*}

\subsection{Connectivity constraint and SWAP operations}

In superconducting quantum systems, a qubit is linked to one or more neighboring qubits via resonators (waveguides) that allow a multi-qubit gate between them. The qubit connectivity graph for an IBM computer is depicted in Fig. \ref{1}). The native 2-qubit gate (CNOT in IBM and CZ in Rigetti) can only be applied between connected qubits. For example, CNOT between qubits 1 and 2 is permitted on the depicted device because there is an edge in the graph between these qubits. Since qubits 1 and 4 are not connected, CNOT cannot be applied directly between them. This limited connectivity is a challenge in quantum circuit mapping (referred to as coupling constraints) which is handled by routing qubits via the SWAP operation so that logical qubits with 2-qubit operations become nearest neighbors, at the cost of:

\textbf{(a) Increase in circuit depth:} extra SWAP operations result in an increase in the overall depth of the quantum circuit. 

\textbf{(b) Computational overhead:} each SWAP operation consumes additional gate resources and prolongs the execution time of the circuit. The increase in gate count can lead to longer computation times, making quantum algorithms more susceptible to noise and decoherence effects.

\textbf{(c) Error accumulation:} as each SWAP operation introduces its own sources of error, extra SWAP operations contribute to error accumulation. The accumulation of errors can severely impact the final results of the quantum computation and reduce the reliability of the output.

\subsection{Relation to prior work}

The field of quantum security has seen significant interest, leading to various research efforts aimed at protecting intellectual property (IP) \cite{b10}\cite{b11}, combating Trojan insertion \cite{b12}\cite{b13}, and dealing with threats from untrusted hardware providers \cite{b14}. The lack of exploration of attack surfaces in a MTC setting, however, is a notable gap in the existing literature. In this work, we aim to bridge this gap and focus on vulnerabilities specific to MTC in superconducting hardware systems. In \cite{b16}, authors propose an adversarial attack model that takes advantage of crosstalk on shared hardware in a MTC setting. 
The adversary's program is designed to exploit crosstalk effects to compromise the victim's quantum computation. They demonstrate the attack's success in lowering a victim's quantum program's output probability. 
In another recent study \cite{b15}, the authors presents a vulnerability in shared trapped-ion (TI) systems that require shuttle operations for communication among traps. They identify the repeated shuttle operations as a critical factor contributing to heightened quantum bit energy and a decline in computation reliability (fidelity). Building upon this insight, the authors propose an attack strategy in which the adversary deliberately designs their program to incur a significant number of shuttles within their own program. This intentional increase in shuttle operations lowers the overall fidelity of the TI system, consequently affecting the fidelity of the victim's program as well. In contrast, our work focuses on superconducting hardware systems, specifically on an attack scenario in which an adversary occupies specific qubits, with the goal of adding extra SWAP operations in the victim circuit during compilation.

\begin{algorithm}[t]
\caption{Quantum Program Scheduler}
\label{algo:scheduler}
\SetAlgoLined
\KwIn{Quantum programs in queue with fair share policy scores.}
\KwOut{Quantum programs with allocated initial qubits for next run.}

Divide the fair-share queue into three priority groups: high, medium, and low, based on scores (lower score $\rightarrow$ higher priority)\;

Calculate the priority metric (PM = total 2-qubit gates / total gates)\;

Initialize an empty queue for selected programs\;

\While{Hardware is running previous batch and programs in the fair-share queue}{
    Select the highest priority program from the queue\;

    \For{Each program in the fair-share queue}{
        \If{Program depth is comparable to the selected program's depth}{
            Check qubit feasibility, connectivity, and mapping for parallel execution\;
            \If{Feasible}{
                Assign the initial qubit mapping based on the program's priority level\;
                    \If{Same priority level}{
                    Assign the initial qubit mapping based on the PM (higher PM $\rightarrow$ higher priority)\;
                    }
                Queue up the program for execution\;
                Update the fair-share queue (remove the selected program)\;
                Break the loop and select the next highest priority program\;
            }
        }
    }
}

\KwOut{Execution order of the selected programs on the quantum hardware}
\end{algorithm}

\section{Scheduler}

The proposed quantum program scheduler algorithm (Algorithm \ref{algo:scheduler}) aims to explore a SWAP-based attack model while ensuring equitable utilization of quantum hardware for optimizing program execution on NISQ computers. The scheduler incorporates the following key features:

\textbf{a) Fair-Share Queuing:} The scheduler uses a fair-share queuing mechanism at its core, dynamically selecting programs for execution to prevent resource monopolization. This approach, like IBM's current policy\cite{b17}, identifies the group with the least utilized share within the scheduling window, ensuring that all users have equal access to quantum hardware. Within each group, program execution is prioritized based on the least used share, with the oldest jobs being prioritized first-in-first-out (FIFO). 

Example 1: To illustrate, we consider a scenario where five jobs are submitted to the hardware and assuming their normalized resource usages (i.e., usage divided by total allocated usage) are as follows: Job1 (0.2), Job2 (0.3), Job3 (0.1), Job4 (0.1), and Job5 (0.3). According to the fair-share queuing, jobs with the least usage will be prioritized, with a FIFO order for older jobs. Therefore, the prioritized order from highest to lowest would be: Job3 (0.1), Job4 (0.1), Job1 (0.2), Job2 (0.3), and Job5 (0.3).

\textbf{b) Priority-Based Grouping:} To enhance fairness and prioritize program execution, the scheduler divides the program queue into three distinct but equal priority groups: high, medium, and low. Programs are assigned priorities based on their scores i.e., lower scores get higher priority. Hence critical programs are given precedence on the quantum hardware.

\textbf{c) Program Selection:} Once priority-based grouping is established, the scheduler selects the program with the highest priority from the fair-share queue for immediate execution. The scheduler then explores opportunities for parallel execution, iteratively checking other programs in the queue with comparable execution time that can be run simultaneously. Thorough checks on qubit requirement, connectivity, mapping, and no qubit sharing are conducted to ensure compatibility and minimize unnecessary SWAP operations. For instance, in Example 1, Job3 is chosen due to its highest priority. The scheduler then evaluates Job4, Job1, Job2, and Job5 iteratively. The selection depends on meeting criteria such as satisfying qubit requirements, having a total qubit count less than the hardware limit, matching execution times, maintaining connectivity, and adhering to compatibility constraints, which ensure continuous qubit connectivity and no sharing between jobs.

\textbf{d) Priority Metric and Resource Allocation:} The scheduler generates a priority metric (PM) for each program by calculating the ratio of \textit{2-qubit gates} to \textit{total gates} in the program. Based on program priorities (high, medium, or low), an initial qubit mapping is assigned. When a tie occurs, the priority metric (the higher the PM, the higher the priority) is used to resolve the mapping decision (for example, if two users request to use the same qubit). The scheduler iterates through programs until the batch running on the hardware is finished or all available qubits for the next run are fully allocated mapping for the next run.

\section{Threat model}


\subsection{Adversary capabilities}

We assume that the adversary has (a) access to the quantum hardware's connectivity graph (typically a public information), detailing permissible two-qubit gate connections and the necessity of SWAP operations between disconnected qubits. 
(b) sufficient computational resources to analyze qubit quality using error rates and the connectivity graph, enabling the selection of suitable qubits to execute an attack.

\subsection{Proposed SWAP injection attack model}

When a quantum job is submitted to a specific quantum system, it enters the scheduler alongside other jobs submitted by different users, forming a pool of tasks awaiting execution on the hardware. The scheduler determines the order of execution for these jobs and which programs can run concurrently on the given quantum hardware for a single run in a MTC environment. An adversary can take advantage of oversights in different schedulers and exploit the connectivity constraints present in the quantum hardware. In this context, both the adversary's program and the victim's program are executed on a quantum hardware with a predefined connectivity graph. The graph defines direct interactions between certain qubits through two-qubit gates (e.g., CNOT gates), while other qubit pairs require SWAP operations to interact indirectly. The adversary's objective is to maximize the number of additional SWAP operations imposed on the victim's program by strategically occupying specific qubits of the connectivity graph. The outlined threat model is depicted in Fig. \ref{2}. To illustrate the impact of this attack model, we present a detailed case study using a sample scheduler. Through strategic adversarial qubit allocation, we demonstrate how the attack induces an increase in SWAP operations on the victim's program compared to an alternate allocation baseline scenario.

\begin{figure}
    \centering
    \includegraphics[width= 3.5in]{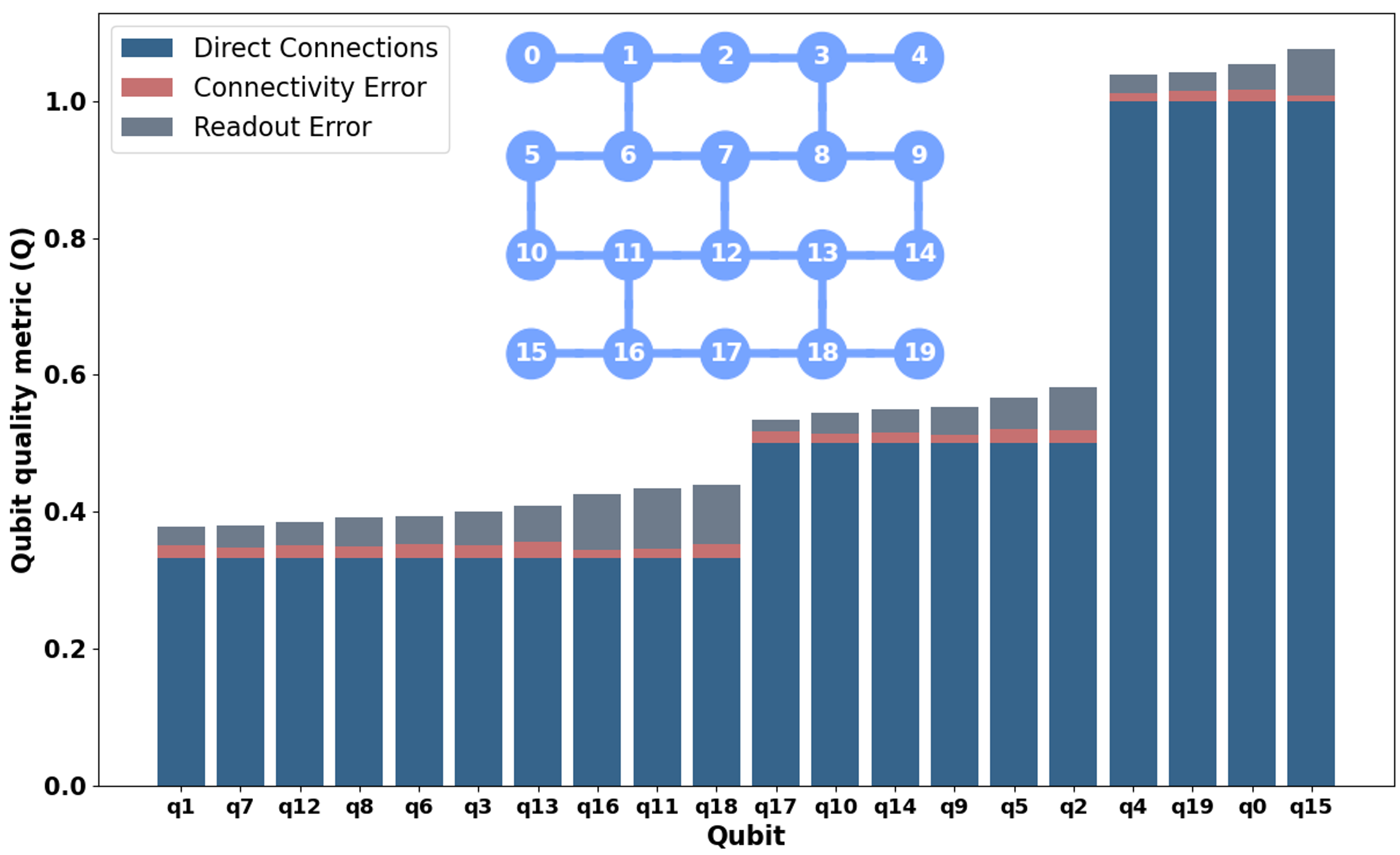}
    \caption{Qubit quality metric (Q) (where {W1,W2,W3}$=0$) for a sample hardware with the shown connectivity graph. The lower the Q, the more likely it will be targeted by an adversary. }
    \label{3}
\vspace{-4mm}
\end{figure}

\subsection{Attack scenario : Kitchen sink approach}

We consider a ``kitchen sink" attack scenario in which an adversary targets victim programs using a variety of adversarial programs. The adversary employs a comprehensive strategy to exploit scheduler vulnerabilities (e.g., fair share allocation policy) despite not knowing the exact target program details (e.g., priority, number of qubits, execution time and job number). The adversary, for example, can hog up certain qubits by submitting multiple high-priority, medium-priority, and low-priority jobs with varying qubit and execution time requirements and high priority metric (Section III.d) biasing the scheduler for qubit allotment thereby limiting victim program access and resulting in unavoidable SWAP operations. This will ensure the victim is attacked regardless of their position and priority in the queue.

\subsection{Adversarial qubit selection}

The choice of qubits for a SWAP-based attack is critical for the adversary to accomplish the attack efficiently. The adversary can consider several factors when selecting the qubits for this attack, including:

\textbf{(a) Number of Direct Connections:} Qubits with more direct connections provide more incentive for the adversary to occupy.

\textbf{(b) Connection Errors:} Another important factor to consider is the severity of connectivity errors associated with each qubit.

\textbf{(c) Readout Assignment Errors:} readout assignment errors are errors that occur during the measurement process, leading to incorrect measurement outcomes. Qubits with higher readout assignment errors are more likely to produce erroneous measurement results, which can be exploited by the adversary.

\subsubsection{Qubit Quality Metric} We define a qubit quality metric that combines the aforementioned factors to aid in the selection of most viable qubits by the adversary.

\begin{equation}
\text{Q} = \text{1/DC} \cdot W_1 + \text{CE} \cdot W_2 + \text{RE} \cdot W_3
\end{equation}

Where:
\textit{DC} represents the number of direct connections, \textit{CE} represents the connectivity errors, \textit{RE} represents the readout assignment errors, and W1, W2, and W3 are weights assigned to each factor. 

\begin{figure}
    \centering
    \includegraphics[width= 3.5in]{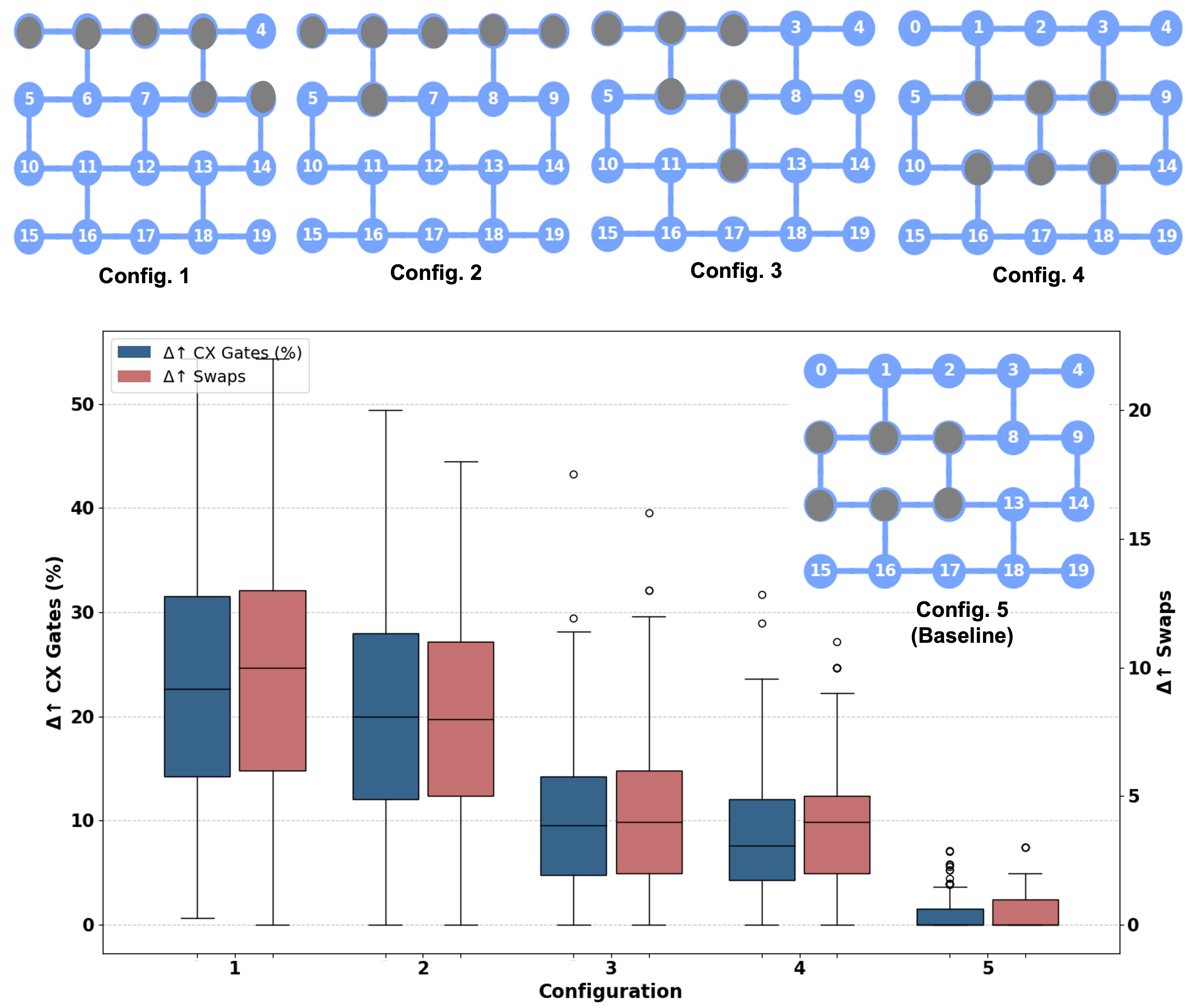}
    \caption{Performance comparison of 100 different 6-qubit programs in five different scenarios of qubit allocation on hardware (fake\_singapore[20 qubit]). We report SWAP overhead (\% increase in CNOT gates) and absolute number of additional SWAPs added, for each configuration compared to configuration 5 (baseline).}
    \label{4}
\vspace{-4mm}
\end{figure}

These weights can be adjusted based on the adversary's goals and the quantum computer's specific characteristics. By calculating the qubit quality metric for each qubit, the adversary can rank the qubits based on their potential to contribute to the success of the SWAP-based attack. For example: Fig. \ref{3} depicts the qubit metric calculated for a backend with (W1, W2, W3) = 1. The higher the qubit quality score, the more likely a qubit will be shortlisted for the attack.

\begin{figure}
    \centering
    \includegraphics[width= 3.5in]{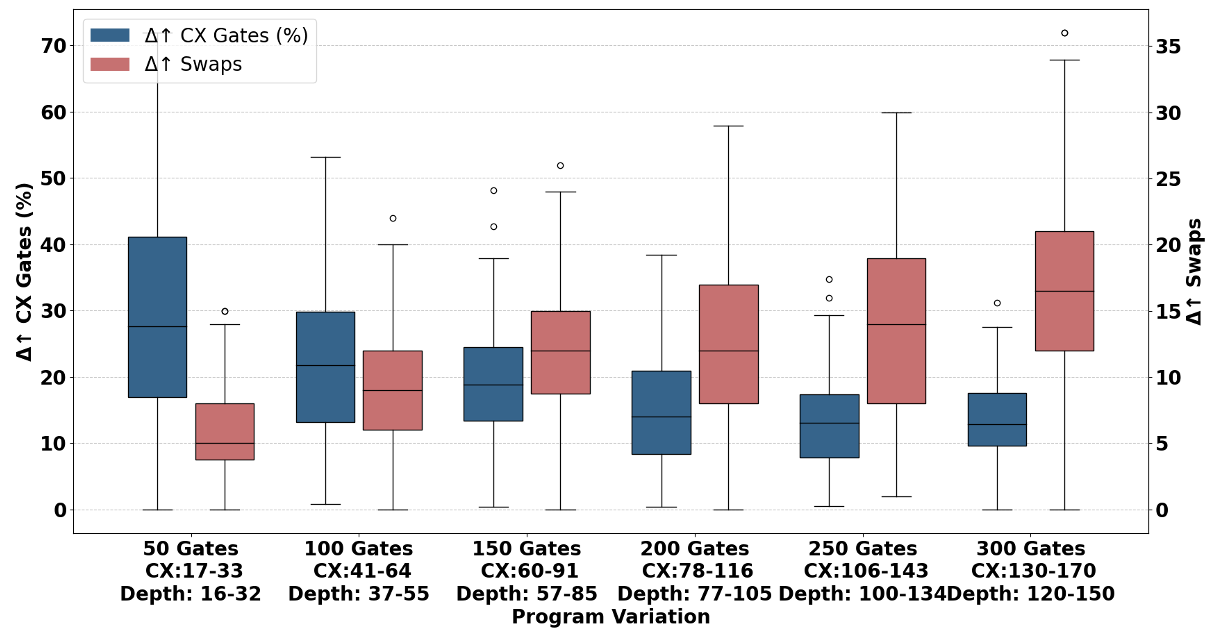}
    \caption{Performance comparison of 100 different 6-qubit program batches with different program parameters (total gates, CNOT(CX) gates and depth) on hardware (fake\_singapore[20 qubit]). We report SWAP overhead (\% increase in CNOT gates) and absolute number of additional SWAPs added, for configuration 1 compared to configuration 5 (baseline).}
    \label{5}
\vspace{-4mm}
\end{figure}

\section{Results and analysis}

\subsubsection{Experimental setup}

We leverage the Qiskit open-source quantum software development kit from IBM, employing a Python wrapper for simulations. For \textbf{benchmarks}, we design random quantum circuits with varying characteristics, including qubit counts, gate counts, circuit depth, and connectivity patterns. For \textbf{benchmark execution}, we utilize Qiskit's fake provider module (fake\_vigo[6 qubit], fake\_singapore[20 qubit]), which comprises noisy simulators mimicking real IBM Quantum systems through system snapshots. These snapshots contain crucial information about the quantum system, such as the coupling map, basis gates, and qubit parameters. During compilation, a SWAP gate is reduced to a combination of CNOT gates (usually 3-CNOT gates). Therefore, we use \textit{SWAP overhead} (\% increase in CNOT gates) and absolute number of additional SWAPs added as our \textbf{performance metric}.

\subsubsection{Attack feasibility: The impact of resource allocation in MTC}

Resource allocation significantly impacts performance of a quantum program in MTC. Long-distance interactions between qubits increase SWAPs, leading to errors and degraded fidelity. Conversely, higher connectivity reduces SWAPs, enhancing efficiency and mitigating errors. Non-preferential resource allocation due to an adversarial attack may lead to extra added SWAPs. We evaluate the performance of 100 different 6-qubit programs in five different scenarios of qubit allocation on hardware (fake\_singapore[20 qubit]) with connectivity graph, as shown in the Fig.\ref{4}. We contrast these allocation scenarios with a baseline allocation, which represents the most densely connected allocation. We report the additional CNOT and SWAP operations introduced compared to the baseline allocation \textit{configuration 5}. We observe a max of $\approx$55\% and a median increase of $\approx$25\% in SWAP overhead between configurations 1 and 5 (baseline). Even for the next best connected allocation we report an increase of max $\approx$15\% and a median increase of $\approx$8\% for 100 different circuits. \textbf{An adversarial attack that prevents the victim from occupying the most densely connected configuration results in a detrimental increase in SWAP overhead.}

\subsubsection{Impact of program complexities}

We report the performance of 100 different 6-qubit program batches with different program parameters (total gates, CNOT(CX) gates, and depth) in Fig. \ref{5}. We compare configuration 1 to configuration 5 (baseline) in terms of SWAP overhead (\% increase in CNOT gates) and absolute number of additional SWAPs added. As program complexity increased with higher depth and gate count, the percentage increase in CNOT gates (SWAP overhead) decreased from a mean of $\approx29$\% (for program batch of 50 gates) to $\approx$15\% (for program batch of 300 gates). This implies that the additional computational burden (SWAP overhead) introduced by SWAP operations became less significant in more complex circuits, however the absolute number of SWAP operations increases (from a mean of $\approx10$\% for the program batch of 50 gates to $\approx$35\% for the program batch of 300 gates) as expected. \textbf{The attack is comparably effective, if not more so, on smaller programs in comparison to larger counterparts.}

\subsubsection{Effect of program size (i.e., number of qubits)}

\begin{figure}
    \centering
    \includegraphics[width= 3.5in]{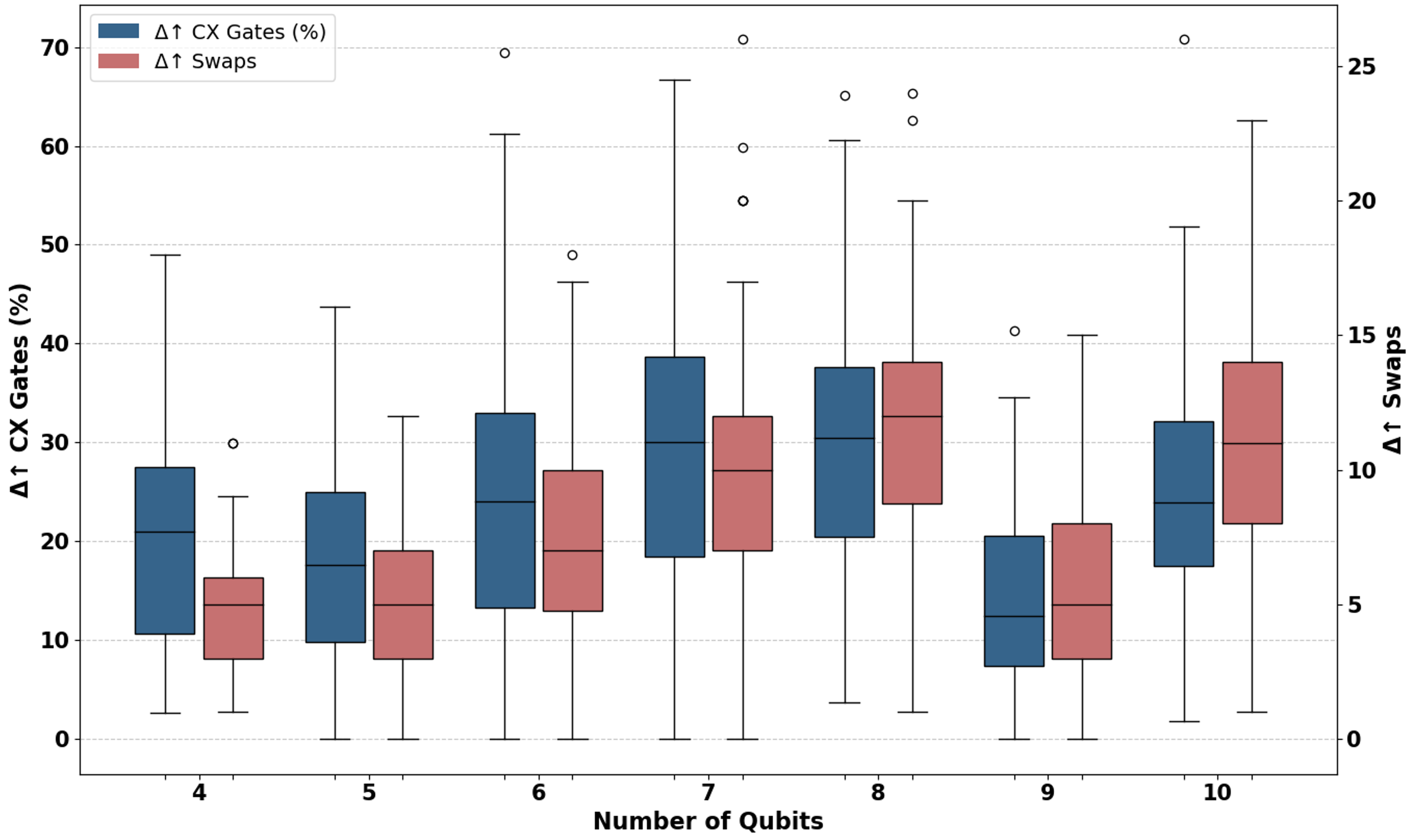}
    \caption{Performance comparison for 100 program batches for each qubit case ranging from 4 to 10 qubits on the Fake\_Singapore with similar depth and gate count. For each qubit case, the least and the most densely connected (baseline) configurations were compared.  }
    \label{6}
\vspace{-4mm}
\end{figure}

We analyzed 100 program batches for each qubit case ranging from 4 to 10 qubits on the Fake\_Singapore. All configurations had similar depth and gate count (Fig. \ref{6}). For each qubit case, the least and the most densely connected (baseline) configurations were compared. The results showed an average increase of $\approx28$\% in mean SWAP overhead, with a maximum increase of $\approx70$\%. \textbf{The extent of SWAP overhead is influenced by both hardware connectivity and the potential density of connections for different qubit programs}.

\section{Defense Strategy}



Schedulers can use anomaly detection to identify unusual submission patterns, such as excessive job requests or skewed prioritization based on factors like job priority, qubit needs, and execution times. When anomalies are detected, alerts can be triggered. To address this, flagged users can either run their programs independently (e.g., in a single execution environment), or schedulers can dynamically adjust priorities to allocate resources more fairly. Machine learning techniques, particularly One-Class SVMs (Support Vector Machines), are well-suited for detecting anomalous user behavior \cite{b18} in program queues, which is essentially a one-class classification problem. We propose a sample model to implement this classifier:

\textbf{Data collection, features and labels:} Schedulers can collect historical data on user behavior in the program queue. This data would include features (with normal and anomalous labels):

\subsubsection{Job request frequency}
\textbf{Normal:} Users who submit up to a certain number of job requests per day.
\textbf{Anomalous:} Users who consistently submit more job requests than the defined threshold.
\subsubsection{Job concurrency} 
\textbf{Normal:} Users who submit a reasonable number of jobs concurrently to the queue in a given time window. 
\textbf{Anomalous:} Users who attempt to run a large number of jobs concurrently.
\subsubsection{Skewed Prioritization} 
\textbf{Normal:} Jobs for a specific user with priority levels distributed evenly across a predefined range.
\textbf{Anomalous:} Jobs for a specific user with priority levels heavily skewed toward either the highest or lowest end of the priority range \subsubsection{User Account Activity}
\textbf{Normal:} Users who maintain a consistent level of account activity.
\textbf{Anomalous:} Users who exhibit bursts of activity or prolonged periods of inactivity.
\subsubsection{Execution Duration vs. Priority}
\textbf{Normal:} Jobs with higher priority generally complete faster than lower-priority jobs.
\textbf{Anomalous:} Jobs with lower priority completing significantly faster than higher-priority jobs.
\subsubsection{Resource Contention}
\textbf{Normal:} Jobs that do not frequently compete for the same resources.
\textbf{Anomalous:} Jobs that frequently contend for the same resources

\textbf{Training and detection:} The One-Class SVM model can be trained exclusively on data representing ``normal" and abnormal behavior. This trained model can subsequently detect anomalies in new data by assigning decision function scores. Data points with scores well below a set threshold are identified as anomalies.


\section{Conclusion}

In this paper 
we introduce a representative scheduler
for multi-tenant computing scenarios and propose a novel SWAP injection attack model. We found up to a $\approx$55\% increase with a median of $\approx$25\% in the swap overhead. We also propose a sample machine learning model to detect any anomalous user behavior as a countermeasure.


\section{Acknowledgment}

This work is supported in parts by NSF (CNS-1722557, CNS-2129675, CCF-2210963, CCF-1718474, OIA-2040667, DGE-1723687, DGE-1821766, and DGE-2113839), Intel's gift and seed grants from Penn State ICDS and Huck Institute of the Life Sciences.


\end{document}